\documentclass[aps, prd, superscriptaddress, nofootinbib,preprintnumbers]{revtex4}
\usepackage{amsmath}
\usepackage{graphicx}
\usepackage{color}

\newcommand{\lesssim}{\mathrel{\mathpalette\vereq<}}

\newcommand{\chushi}[1]{}
\newcommand{\BL}{{B\mathchar`-L}}

\begin{document}

\title{Scale generation via dynamically induced multiple seesaw mechanisms}      
\author{Hiroyuki Ishida}\thanks{{\tt hiroyuki403@cts.nthu.edu.tw}}
	\affiliation{ Physics Division, National Center for Theoretical Sciences, Hsinchu 30013, Taiwan.}
\author{Shinya Matsuzaki}\thanks{{\tt synya@hken.phys.nagoya-u.ac.jp}}
      \affiliation{ Institute for Advanced Research, Nagoya University, Nagoya 464-8602, Japan.}
      \affiliation{ Department of Physics, Nagoya University, Nagoya 464-8602, Japan.}   
\author{Shohei Okawa}\thanks{{\tt okawa@eken.phys.nagoya-u.ac.jp}}
      \affiliation{ Department of Physics, Nagoya University, Nagoya 464-8602, Japan.}
\author{Yuji Omura}\thanks{{\tt yujiomur@kmi.nagoya-u.ac.jp}}
      \affiliation{ Kobayashi-Maskawa Institute for the Origin of Particles and the
Universe, \\ Nagoya University, Nagoya 464-8602, Japan.}

\date{January 3, 2017}

\begin{abstract}
We propose a model which accounts for the dynamical origin of the electroweak symmetry breaking (EWSB), 
directly linking to the mass generation of dark matter (DM) candidate and active neutrinos. 
The standard model (SM) is weakly charged under 
the $U(1)_{\BL}$ gauge symmetry, in conjunction with 
newly introduced three right-handed Majorana neutrinos and the $U(1)_{\BL}$ Higgs. 
The model is built on the classical scale invariance, 
that is dynamically broken by 
a new strongly coupled sector, what is called the hypercolor (HC) sector, 
which is also weakly coupled to the $U(1)_{\BL}$ gauge.    
At the HC strong scale, 
the simultaneous breaking of 
the EW and $U(1)_{\BL}$ gauge symmetries is triggered 
by dynamically induced multiple seesaw mechanisms, namely bosonic seesaw mechanisms. 
Thus, all the origins of masses are provided singly by the HC dynamics:      
that is what we call the {\it dynamical scalegenesis}. 
We also find that an HC baryon, with mass on the order of a few TeV, 
 can be stabilized by the HC baryon number 
and the $U(1)_{\BL}$ charge, so identified as a DM candidate.    
The relic abundance of the HC-baryon DM 
can be produced 
dominantly via the bosonic-seesaw portal process, and the HC-baryon DM can be measured 
through the large magnetic moment coupling generated from the HC dynamics, or 
the $U(1)_{\BL}$-gauge boson portal in direct detection experiments.  
\end{abstract}
\maketitle

\section{Introduction}\label{Sec1}

The standard model (SM) of particle physics has achieved great success
and been excellently consistent with experiments so far.  
In the SM, an elementary scalar field, a Higgs field, plays a role in spontaneously breaking electroweak symmetry 
and generating masses, and the signals predicted 
by the SM Higgs boson have been discovered at the LHC~\cite{Aad:2012tfa,Chatrchyan:2013lba}. 
However, the source to trigger the electroweak symmetry breaking (EWSB) seems quite ad hoc and mysterious: 
one needs to assume square of the Higgs mass parameter to be negative without any dynamical reason. 
In that sense, the mechanism of the EWSB in the SM is still unsatisfactory, so one is urged to go beyond 
the SM, including new physics, where the low-energy physics looks much like that of the SM.

Once going beyond the SM, 
to reveal the origin of the EWSB, triggered by the negative mass squared for the Higgs, 
one necessarily encounters a problem: 
cancellation of quantum corrections to the Higgs mass which is proportional to the new physics scale. 
One way to avoid this problem is to invoke the classical scale invariance, 
which can forbid all dimensional parameters including the Higgs mass in the theory, 
hence one is to be free from quantum corrections to 
the Higgs mass~\footnote{
Note that the scale symmetry is 
anomalous to be explicitly broken by quantum corrections, 
yielding the trace anomaly. 
The gravitational effects may, however, cancel 
the trace anomaly and make the theory  
asymptotically safety~\cite{Shaposhnikov:2009pv,Helmboldt:2016mpi}. 
Therefore, we assume that the classical scale invariance is held below
the Planck scale, as long as all the couplings in the theory do not reach  
the Landau pole up to the Planck scale, as argued in~\cite{Helmboldt:2016mpi}. 
\label{footnote:safety}}. 
To retrieve the EWSB, one thus needs to generate the nonvanishing and negative squared Higgs mass term somehow.

One idea to generate the Higgs mass in the scale-invariant models 
is to introduce $U(1)_{\BL}$ gauge symmetry, which would be inspired by the possible existence of grand unified theory.  
In this scenario, 
the $U(1)_{\BL}$ symmetry is broken by the newly introduced the vacuum expectation value of the 
$U(1)_{\BL}$ Higgs boson generated by radiative corrections, 
so-called Coleman-Weinberg mechanism~\cite{Coleman:1973jx}. 
Then the mass term of Higgs is induced via the mixing term between the SM Higgs and the $U(1)_{\BL}$ Higgs bosons~\cite{B-L}.

Another benefit to introduce the $U(1)_{\BL}$ gauge symmetry involves physics related to neutrinos and 
their mass generation mechanism. 
When the $U(1)_{\BL}$ gauge symmetry is encoded into the classical scale invariant scenario, the neutrino mass generation 
is achieved by nonzero vacuum expectation value of $U(1)_{\BL}$ Higgs, where the neutrinos possess the right-handed (RH) Majorana nature.    
Thus, the extension by the $U(1)_{\BL}$ gauge symmetry can explain the origin 
of both the EWSB and Majorana mass for RH neutrinos simultaneously~\cite{nu:sumaary}.

In the scenario of this class, however, the mixing coupling between the EW Higgs and $U(1)_{\BL}$ Higgs fields  
is needed to be tuned to be negative to make the square of Higgs mass negative, 
i.e., to realize the EWSB, so this idea seems to be still unsatisfactory~\footnote{
When the EW Higgs sector is extended from the minimal structure, 
the negative mass term can be generated without assuming ad hoc negative quartic coupling mixing~\cite{Okada:2014nea}.}.

Another proposal built on the classical scale invariance has been published 
in the framework of namely the bosonic seesaw mechanism~\cite{Haba:2015lka,BSS}, which is triggered by   
a new strong dynamics~\cite{Haba:2015qbz,Ishida:2016ogu,Ishida:2016fbp}, what we call the hypercolor (HC).  
In models of this class, the scale-invariance is dynamically broken by the strong scale intrinsic 
to the HC dynamics, and the negative-mass squared of the Higgs 
is then dynamically generated by the seesaw mechanism operative 
between the elementary Higgs field and a composite Higgs field generated from the HC dynamics. 
(Since the sign is never absorbed by phase rotations in the case of boson fields, 
the negative sign induced by the seesaw mechanism is manifestly physical to be a trigger of the EWSB.)  
\footnote{The idea of the scale generation by dimensional transmutation 
from a hidden strong dynamics, or the existing QCD, has been discussed in the literature~\cite{Hid-strong-EWSB} 
in a context different from the present model based on the bosonic seesaw mechanism. 
}

In this paper, we develop the bosonic seesaw model including the $U(1)_{\BL}$ gauge symmetry: 
all the masses, for the SM particles and $U(1)_{\BL}$ Higgs, gauge boson, and right-handed Majorana neutrinos (RHM$\nu$s), 
are generated singly by the new strong dynamics, the HC, via a sequence of bosonic seesaws (multiple seesaws) 
involving the HC composite Higgs bosons:  
that is the {\it dynamical scalegenesis}. 
The scale of active neutrino masses is generated via the neutrino seesaw of 
ordinary type-I form~\cite{seesaw}, which is induced from 
the bosonic seesaw term of the elementary and composite $U(1)_{\BL}$ Higgs bosons.

We also find that the lightest $U(1)_{\BL}$-charged HC baryon can be a dark matter (DM) candidate, 
and that the relic abundance can nonthermally be produced  
via the bosonic-seesaw portal process to explain the observed amount.  
The HC-baryon DM possesses the sensitivity enough to be accessible in direct detection experiments, 
due to the large magnetic-moment $g$ factor generated by the strong HC dynamics, 
or the sizable $U(1)_{\BL}$ gauge boson portal coupling.

This paper is organized as follows:
In Sec.\ref{Sec2} we introduce our model and show how the {\it dynamical scalegenesis} works in Sec.\ref{Sec3}. 
We discuss our dark matter candidate in Sec.\ref{Sec4}, and finally our conclusion is given in Sec.\ref{Sec5}.
The Appendix \ref{App1} compensates the potential analysis to realize the EWSB and $U(1)_{\BL}$ breaking.

\section{Model}\label{Sec2} 
Our model consists of the HC sector having the $SU(3)_{\rm HC}$ gauge symmetry 
and the SM sector.  
The key assumption in this model is presence of the classical scale invariance,  
so that the Higgs field $(H)$ in the SM sector does not have the mass term. 
The HC sector includes eight HC gluons (${\cal G}$) of the $SU(3)_{\rm HC}$ as well as 
four HC fermions $(F_{i=1,2,3,4})$ forming the the fundamental representation of $SU(4)$,  
$F_{L/R}=(\chi,\psi_1,\psi_2)^T_{L/R}$. 
The HC dynamical feature is assumed to be a complete analogue of QCD. 

In addition to the SM gauge symmetry, we introduce the $\BL$ gauge symmetry, $U(1)_{\BL}$, 
by which the $U(1)_{\BL}$ gauge boson ($X$) and  
a new complex scalar ($\phi$) weakly couple involving the HC sector and the SM particles.  
The HC fermions are vector-likely charged under the SM and $U(1)_{\BL}$ gauges (see Table~\ref{tab:1}).  
To make the $U(1)_{\BL}$-gauge anomaly-free,    
we also introduce right-handed Majorana neutrinos (RHM$\nu$s) $N_R^{1,2,3}$.

\begin{table}[t]
\vspace{5mm}
 \begin{tabular}{c||c|c|c|c|c}
\hline 
$F_{L/R}$ & $SU(3)_{\rm HC}$ & $SU(3)_c$ & $SU(2)_W$ & $U(1)_Y$ & $U(1)_{\BL}$ \\ 
\hline 
$\chi =(\chi_1,\chi_2)^T $ & 3 & 1 & 2 & $1/2 + q$ & $q'$   \\ 
$\psi_1$ & 3 & 1 & 1 & $q$ & $q'$ \\ 
$\psi_2 $ & 3 & 1 & 1 & $q$ & $-2 + q'$  \\ 
\hline 
\end{tabular} 
\caption{The charge assignment for the HC fermions under the HC ($SU(3)_{\rm HC}$), SM gauges ($SU(3)_c \times SU(2)_W \times U(1)_Y$) 
and $U(1)_{\BL}$ gauge, where $q$ and $q'$ are arbitrary numbers.} 
\label{tab:1} 
\end{table} 
\begin{table}[t]
 \begin{tabular}{c|c|c|c|c}
\hline 
 & $SU(3)_c$ & $SU(2)_W$ & $U(1)_Y$ & $U(1)_{\BL}$ \\ 
\hline 
$q_L^\alpha$ & 3 & 2 & $1/6$ & $-1/3$   \\ 
$u_R^\alpha$ & 3 & 1 & $2/3$ & $-1/3$ \\
$d_R^\alpha$ & 3 & 1 & $-1/3$ & $-1/3$ \\ 
$l_L^\alpha$ & 1 & 2 & $-1/2$ & 1 \\
$e_R^\alpha$ & 1 & 1 & $-1$ & 1 \\
$H$ &  1 & 2 & $1/2$ & 0 \\  
\hline 
$N_R^\alpha$ & 1 & 1 & 0 & 1 \\ 
$\phi$ & 1 & 1 & 0 & $-2$   \\ 
\hline 
\end{tabular} 
\caption{The charge assignment for the SM quarks ($q_L^\alpha, u_R^\alpha, d_R^\alpha$), leptons ($l_L^\alpha, e_R^\alpha$),  
Higgs ($H$) and three Majorana neutrinos $(N_R^\alpha)$ and 
a $U(1)_{\BL}$ complex scalar ($\phi$). 
The upper script $\alpha$ attaching on fermion fields denote the generation index, $\alpha = 1,2,3$. 
All the fields listed here do not carry the $SU(3)_{\rm HC}$ charge.} 
\label{tab:2}
\end{table}

Regarding the matter contents of the model, the charge assignment for these gauges 
is summarized in Tables~\ref{tab:1} and ~\ref{tab:2}.  
Reflecting the gauge symmetries read off from Table~\ref{tab:1} and~\ref{tab:2} 
and the classical scale-invariance,  
one can uniquely write down the model Lagrangian:   
\begin{eqnarray} 
{\cal L} &=& 
{\cal L}^{\rm gauge-kin.} 
+ \overline{q_L^\alpha} i \gamma_\mu D^\mu q_L^\alpha 
+ \overline{q_R^\alpha} i \gamma_\mu D^\mu q_R^\alpha 
+ \overline{l_L^\alpha} i \gamma_\mu D^\mu l_L^\alpha 
+ \overline{e_R^\alpha} i \gamma_\mu D^\mu e_R^\alpha 
+ |D_\mu H|^2 
+ |D_\mu \phi|^2 
+ 
\nonumber \\ 
&& 
+ \overline{N_R^\alpha} i \gamma_\mu D^\mu N_R^\alpha 
+ \overline{F} i \gamma_\mu D^\mu F 
- \left( y^{\alpha\beta}_u \overline{q_L^\alpha} \tilde{H} u_R^\beta 
+ y^{\alpha\beta}_d \overline{q_L^\alpha} H d_R^\beta 
+ y^{\alpha\beta}_e \overline{l_L^\alpha} H e_R^\beta 
+ y^{\alpha\beta}_{lN} \overline{l_L^\alpha} \tilde{H} N_R^\beta
 + {\rm h.c.} \right) 
\nonumber \\ 
&& 
+ y_H (\overline{\chi} H \psi_1 + {\rm h.c.} ) 
+ y_\phi (\overline{\psi_2} \phi \psi_1 + {\rm h.c.}) 
+ y_N^{\alpha \alpha} (\phi \overline{N^{c \alpha}_R} N_R^\alpha + {\rm h.c.}) 
- \lambda_H (H^\dag H)^2 - \lambda_\phi (|\phi|^2)^2 
- \kappa_\phi |\phi|^2 (H^\dag H) 
\,, \label{Lag}
\end{eqnarray}
where ${\cal L}^{\rm gauge-kin.}$ stands for the kinetic terms of all gauge fields forming the gauge field strengths;    
the covariant derivatives ($D_\mu$) can be read off from Tables~\ref{tab:1} and~\ref{tab:2}; 
the sums over repeated flavor 
indices $\alpha$ and $\beta$ have been taken into account; 
we have chosen 
basis for the $N_R$-flavor structure to be diagonal in the $y_N$-Yukawa coupling. 
The Yukawa couplings $y_H$ and $y_\phi$ are assumed to be much smaller than ${\cal O}(1)$, $y_H \ll 1$ and $y_\phi \ll 1$, 
which will be consistent with realization of the EWSB and $U(1)_{\BL}$ breaking, as will be seen later on.

It is the HC dynamics that generates all the mass scales for the model particles: 
as in QCD, the HC gauge coupling gets strong to dynamically break 
the scale-invariance by the intrinsic scale $\Lambda_{\rm HC}$, say, 
${\cal O}(5-10)$ TeV, 
as the consequence of the dimensional transmutation. 
As will turn out, the HC dynamics triggers a sequence of seesaw mechanism (multiple seesaws) 
so that the dynamically generated scale $\Lambda_{\rm HC}$ 
drives the EWSB as well as 
the mass generation of the active neutrinos.

\section{Bosonic seesaws: {\it dynamical scalegenesis}}\label{Sec3}

The HC sector possesses the (approximate) global ``chiral'' 
$SU(4)_{F_L} \times SU(4)_{F_R}$ symmetry, which is explicitly 
broken by the gauges as seen from Table~\ref{tab:1}, and a couple of Yukawa terms as displayed in 
Eq.(\ref{Lag}).   
At the strong scale $\Lambda_{\rm HC}$,   
this approximate ``chiral'' symmetry is   
spontaneously broken down to the vectorial $SU(4)_{F_V}$ 
by developing the nonzero ``chiral" condensate $\langle \bar{F}^iF^j 
\rangle \sim \Lambda_{\rm HC}^3 \delta^{ij}$, 
to give rise to the  
HC fermion dynamical mass of ${\cal O}(\Lambda_{\rm HC})$.

\begin{table}
\begin{tabular}{c|c|c|c|c}
\hline
${\cal M} = {\cal S} + i {\cal P}$
& constituent &$SU(2)_W$ & $U(1)_Y$ & $U(1)_{\BL}$ \\
\hline
$(f^{\rm HC}_0, a^{\rm HC}_0 + i {\cal P}_{a^{\rm HC}_0})_{ij}$
& $\overline{\chi_i}$ $\chi_j$ & (1, 3) & 0 & 0 \\
$(\Theta_1 + i {\cal P}_{\Theta_1})_i$ & $\overline{\psi_1}$ $\chi_i$ & 2 & $1/2$ & 0 \\
$(\Theta_2 + i {\cal P}_{\Theta_2})_i$ & $\overline{\psi_2}$ $\chi_i$ & 2 & $1/2$ & 2 \\
$\Phi + i {\cal P}_\Phi$ & $\overline{\psi_1}$ $\psi_2$ & 1 & 0 & $-2$ \\
$\varphi_1 + i {\cal P}_{\varphi_1}$ & $\overline{\psi_1}$ $\psi_1$ & 1 & 0 & 0 \\
$\varphi_2 + i {\cal P}_{\varphi_2}$ & $\overline{\psi_2}$ $\psi_2$ & 1 & 0 & 0 \\ 
\hline
\end{tabular}
\caption{The list of the composite scalars and pseudoscalars
embedded in the ``chiral'' $SU(4)_{F}$-16 plet.
In the second row the isosinglet pNGB (like $\eta'$ in QCD)
component has been discarded. }
\label{tab:3}
\end{table}

At the scale $\Lambda_{\rm HC}$, 
the HC sector dynamics can be described as the ``hadron" physics (HC hadron). 
As in the case of QCD, 
the lightest HC hadron spectra are then expected to be constructed from 
the composite scalars $({\cal S})$ and pseudoscalars $({\cal P})$, 
pseudo Nambu-Goldsone bosons (pNGBs) associated with the spontaneous``chiral" symmetry breaking~\footnote{
The mass generation of the pNGBs will be subject to the presence of an extra elementary 
pseudoscalar as discussed in the literature~\cite{Ishida:2016ogu}, 
where the size of masses can be fixed by other couplings irrespectively to those 
presented in Eq.(\ref{Lag}).}.  
The composite scalars acquire the masses of ${\cal O}(\Lambda_{\rm HC})$ 
due to the ``chiral'' symmetry breaking. 
The light HC hadrons form the chiral $SU(4)_{F_L} \times SU(4)_{F_R}$ 
16-plet, ${\cal M} = {\cal S} + i {\cal P}$, 
which can be classified with respect to 
the weak isospin ($SU(2)_W$), $U(1)_Y$ and $U(1)_{\BL}$ charges. 
The complete list for the lightest composite scalars and pseudoscalars 
is provided in Table~\ref{tab:3}.

Besides the HC scalars and pseudoscalars, 
the HC baryons formed by HC fermions like $\sim  F F  F $  
are expected to be generated to have the masses of ${\cal O}(\Lambda_{\rm HC})$ in a way analogously to QCD.  
According to the QCD baryon spectroscopy, 
the spin 1/2 baryons form the $SU(4)_{F_V}$-20 plet, classified by 
the weak isospin ($SU(2)_W$) and $U(1)_Y$ charges. 
In Table~\ref{tab:baryons} the spin 1/2 HC baryons are listed.

Among those HC hadrons,  
we note composite scalars,  
\begin{equation} 
\Theta_1 \sim \overline{\psi_1} \chi 
\,, \qquad 
\Phi \sim \overline{\psi_1} \psi_2 
\,, 
\end{equation} 
in which the $\Theta_1$ has the same quantum numbers  
as those of the elementary Higgs doublet $H$, 
and the $\Phi$ carries the same charges as those the elementary $U(1)_{\BL}$ scalar $\phi$ does. (See Table~\ref{tab:3}).  
Of interest is to note that at the $\Lambda_{\rm HC}$ scale, 
the Yukawa terms with the couplings $y_H$ and $y_\phi$ in Eq.(\ref{Lag}) 
induce the mixing between $\Theta_1$-$H$ and $\Phi$-$\phi$, such as 
$y_H \Lambda_{\rm HC}^2 (\Theta_1^\dag H + {\rm h.c.})$ and $y_\phi \Lambda^2_{\rm HC} 
(\Phi^* \phi + {\rm h.c})$. 
Thus, the mass matrices of the seesaw form are generated:  
\begin{equation} 
 \left(
\begin{array}{cc} 
0 & y_{H/\phi} \Lambda^2_{\rm HC} \\ 
y_{H/\phi} \Lambda^2_{\rm HC} & \Lambda_{\rm HC}^2  
\end{array}
\right)
\,. \label{phi:seesaw}
\end{equation}
These matrices yield the negative mass-squared for $H$ and $\phi$, 
$m_H^2 \simeq - y_H^2 \Lambda_{\rm HC}^2 $ 
and 
$m_\phi^2 \simeq - y_\phi^2 \Lambda_{\rm HC}^2$ for small Yukawa couplings. 
Combined with the quartic potential ($\lambda_H$ and $\lambda_\phi$) 
terms for $H$ and $\phi$ in Eq.(\ref{Lag}), the EWSB and $U(1)_{\BL}$ breaking 
are thus triggered to develop the vacuum expectation values (VEVs) $v_{\rm EW} \simeq 246$ GeV 
and $v_{\phi_1} = {\cal O}(\Lambda_{\rm HC})={\cal O}(5-10\,{\rm TeV})$. 
Then, the physical Higgs boson $(h_1)$ and the $U(1)_{\BL}$ Higgs boson $(\phi_1)$ respectively 
arise around the VEVs $v_{\rm EW}$ and $v_{\phi_1}$, obtaining the masses  
$m_{h_1} \simeq \sqrt{2 \lambda_H} v_{\rm EW} \simeq 125$ GeV and 
$m_{\phi_1} \simeq  2\sqrt{2 \lambda_\phi} v_{\phi_1} \simeq {\cal O}(10 - 30)$ 
TeV  for $\lambda_\phi = {\cal O}(1)$.   
(The detailed potential analysis is given in Appendix~\ref{appendix:2}.)

\begin{table} 
 \begin{tabular}{c|c|c|c|c|c} 
\hline 
HC baryon & constituent & $SU(2)_W$ & 
$I_3$ & $Y$ & 
 $\BL$ \\ 
\hline 
$p_{\rm HC}^{2+3q}$ & $\chi_1 \chi_1 \chi_2$ & 2 & $1/2$ & $3/2 + 3q$ &  $3q'$ \\ 
$n_{\rm HC}^{1+3q}$ & $\chi_1 \chi_2 \chi_2$ & 2 & $-1/2$ & $3/2 + 3q$ &  $3q'$ \\ 
\hline 
$\Lambda_{(1)}^{1+3q}$ & $\chi_1 \chi_2 \psi_1$ & 1 & 0 & $1+3q$ &  $3q'$ \\ 
\hline 
$\Sigma_{(1)}^{2+3q}$ & $\chi_1 \chi_1 \psi_1$ & 3 & 1 & $1+3q$ &  $3q'$ \\  
 $\Sigma_{(1)}^{1+3q}$ & $\chi_1 \chi_2 \psi_1$ & 3 & 0 & $1+3q$ &  $3q'$ \\ 
  $\Sigma_{(1)}^{3q}$ & $\chi_2 \chi_2 \psi_1$ & 3 & $-1$ & $1+3q$ & $3q'$ \\ 
\hline 
$\Xi_{(11)}^{1+3q}$ & $\chi_1 \psi_1 \psi_1$ & 2 & $1/2$ & $1/2 +3q$  & $3q'$ \\ 
 $\Xi_{(11)}^{3q}$ & $\chi_2 \psi_1 \psi_1$ &2 & $-1/2$ & $1/2 +3q$ & $3q'$ \\ 
\hline \hline 
$\Lambda_{(2)}^{1+3q}$ & $\chi_1 \chi_2 \psi_2$ & 1 & 0 & $1 + 3q$ & $-2 + 3q'$ \\ 
 $\Omega_{(12)}^{3q}$ & $\psi_1\psi_1\psi_2$  & 1 & 0 & $3q$ & $-2 + 3q'$ \\ 
\hline 
$\Sigma_{(2)}^{2+3q}$ & $\chi_1 \chi_1 \psi_2$ &3 & 1 & $1 + 3q$  & $-2 + 3q'$ \\  
 $\Sigma_{(2)}^{1+3q}$ & $\chi_1 \chi_2 \psi_2$ &3 & 0 & $1 + 3q$ & $-2 + 3q'$ \\ 
  $\Sigma_{(2)}^{3q}$ & $\chi_2 \chi_2 \psi_2$ &3 & $-1$ & $1 + 3q$ & $-2 + 3q'$ \\ 
\hline 
$\Xi_{(12)}^{1+3q}$ & $\chi_1 \psi_1 \psi_2$ & 2 & $1/2$ & $1/2 + 3q$ & $-2 + 3q'$ \\ 
 $\Xi_{(12)}^{3q}$ & $\chi_2 \psi_1 \psi_2$ & 2 & $-1/2$ & $1/2 + 3q$ & $-2 + 3q'$ \\ 
\hline \hline 
$\Omega_{(22)}^{3q}$ & $\psi_1\psi_2\psi_2$ & 1 & 0 & $3q$ & $-4 + 3q'$ \\ 
\hline 
$\Xi_{(22)}^{1+3q}$ & $\chi_1 \psi_2 \psi_2$ & 2 & $1/2$ & $1/2 + 3q$ & $-4 + 3q'$ \\ 
 $\Xi_{(22)}^{3q}$ & $\chi_2 \psi_2 \psi_2$ &2 & $-1/2$ & $1/2 + 3q$ & $-4 + 3q'$ \\ 
\hline
 \end{tabular} 
\caption{The list of the HC baryons with spin 1/2 
forming the $SU(4)_{F}$-20 plet classified by 
the weak isospin and hypercharge as well as the $U(1)_{\BL}$ charge. 
The upper script on the HC baryons denotes the electromagnetic charge $(Q_{\rm em}=I_3 + Y)$. 
In the list the isospin doublet $\Xi_{12}$ includes 
two degenerate states, analogously to the $\Xi_c,\Xi_c'$
baryons predicted in the quark model applied to QCD.}
\label{tab:baryons}
\end{table}

The $U(1)_{\BL}$-gauge breaking VEV, $v_{\phi_1}$, makes the 
$U(1)_{\BL}$ gauge boson ($X$)  
and RHM$\nu$ $N_R^\alpha$ massive as well: 
by the $\phi$-Higgs mechanism through the covariantized kinetic term $|D_\mu \phi|^2$ in Eq.(\ref{Lag}), 
the $U(1)_{\BL}$ gauge boson gets the mass of order 
${\cal O}(g_{X} \Lambda_{\rm HC}) = {\cal O}(5-10\,{\rm TeV})$ 
with the $U(1)_{\BL}$ gauge coupling of ${\cal O}(1)$; 
the $N_R^\alpha$ become  
massive via the Yukawa coupling $y^{\alpha\alpha}_N$ in Eq.(\ref{Lag}), to get 
the masses $m_{N_R}^{\alpha\alpha} = {\cal O}(y^{\alpha\alpha}_N \Lambda_{\rm HC}) 
= {\cal O}(5-10\,{\rm TeV})$ with $y_N^{\alpha\alpha}={\cal O}(1)$.  
We then note that the $N_R$-mass generation combined with 
the $y_{lN}$-Dirac Yukawa term in Eq.(\ref{Lag}) 
induces the neutrino seesaw: 
\begin{equation} 
\left( 
 \begin{array}{cc}
 0 &  y_{lN}  v_{\rm EW} \\ 
y_{lN}^T  v_{\rm EW} 
 & m_{N_R} 
\end{array}
\right) 
\,. \label{nu:seesaw}
\end{equation}
One can realize the neutrino mass scale    
$m_\nu \simeq y_{lN}^2 v_{\rm EW}^2/m_{N_R}    
={\cal O}(0.1\,{\rm eV})$ for $y_{lN} = {\cal O}(10^{-5})$.

Thus, the HC dynamics triggers the sequence of the bosonic seesaws, 
to generate the masses of all the particles involving 
the SM contents together with a couple of new particles involving a number of HC hadrons, 
the $U(1)_{\BL}$ $\phi$-Higgs boson, gauge boson and 
heavy RHM$\nu$.

\section{Dark matter}\label{Sec4}

The HC baryons possess the HC baryon number associated with the 
global $U(1)_{F_V}$ symmetry, so can be stabilized to be DM candidates.  
Looking at Table~\ref{tab:baryons}, one may expect 
the isosinglets are favored to be the candidates, 
i.e., $\Lambda_{(1),(2)}^{1+3q}$ or $\Omega_{(12),(22)}^{3q}$. 
Since the DM has to be electromagnetically neutral, 
below we shall employ the possible two cases with   
(I) $q=-1/3$ and (II) $q=0$~\footnote{
The value of $q$ would be sensitive to realization of 
the asymptotic safety condition (i.e. no Landau pole up to the Planck scale)  
as noted in footnote~\ref{footnote:safety}. 
We find that the gauge coupling of U(1)$_Y$ does not diverge
below the Planck scale at the one-loop level, as far as $-1 < q < 1/2$ is satisfied.  
Including the two-loop corrections, the bound could be relaxed because of the corrections
from the Yukawa couplings.
\label{footnote:q}}, 
and discuss the stability of the DM candidates, 
the thermal history and the discovery sensitivity in direct detection experiments~\footnote{
The charge $q$ could take arbitrary fractional numbers (satisfying the asymptotic safety condition 
in footnote~\ref{footnote:q})
so that some HC baryons other than those in Cases I and II could be stable. 
In the present study we will disregard this possibility for simplicity. }.

\subsection{Case I with $q=-1/3$}

First, in this case, 
the electromagnetically charged HC bayons in the isospin multiplets 
decay to the neutral-isospin partners along with the $W$ boson emission, 
such as 
\begin{eqnarray}
&& 
p_{\rm HC}^+ \to n_{\rm HC}^0 + W^{+ (*)}
\,, \nonumber \\ 
&& 
\Sigma_{(1),(2)}^\pm \to \Sigma_{(1),(2)}^0 + W^{\pm (*)}
\,, \nonumber \\  
&& 
\Xi_{(11),(12),(22)}^- \to \Xi_{(11),(12),(22)}^0 + W^{-(*)}\,, 
\end{eqnarray}  
where the charged HC baryons have masses larger than the neutral ones 
by the size of ${\cal O}(\alpha_{\rm em} \Lambda_{\rm HC})$ 
(= ${\cal O}(100\,{\rm GeV})$ for $\Lambda_{\rm HC}={\cal O}(5-10\,{\rm TeV})$). 
Then, these neutral HC baryons decay to the SM-singlet $\Lambda_{(1)}^0$ 
or $\Lambda_{(2)}^0$ by emitting the various (off-shell) neutral HC pions listed 
in Table~\ref{tab:3}, which finally decay to diphoton: 
\begin{eqnarray} 
&& 
n_{\rm HC}^0 \to {\cal P}_{\Theta_2}^{0} + \Lambda_{(2)}^0
\,, \nonumber \\ 
&& 
\Sigma_{(1),(2)}^0 \to {\cal P}_{a_0^{\rm HC}}^0 + \Lambda_{(1),(2)}^0
\,, \nonumber \\  
&& 
\Xi_{(11),(12),(22)}^0 \to \widetilde{\cal P}_{\Theta_1, \Theta_1, \Theta_2}^{0} + \Lambda_{(1),(2),(2)}^0
\,, 
\end{eqnarray}  
where the masses of the parent HC baryons are larger than those of the daughters 
by amount of ${\cal O}(\alpha_{\rm em} \Lambda_{\rm HC})$ due to 
the weak interaction corrections~\footnote{
Here the decays of $\widetilde{\cal P}_{\Theta_1}^{0}$ involve the $\psi_1 \leftrightarrow \chi_{1,2}$ conversion 
via the $y_H$-Yukawa coupling in Eq.(\ref{Lag}) with the $H$-Higgs VEV $v_{\rm EW}$, 
while those of $\widetilde{\cal P}_{\Theta_2}^{0}$ 
do the $\phi_1 \leftrightarrow \phi_2$ conversion, as well as the $\phi_1 \leftrightarrow \chi_{1,2}$ conversion arising from 
the $y_\phi$-Yukawa coupling in Eq.(\ref{Lag}) with the $\phi$-Higgs VEV $v_{\phi_1}$.}.

Second, 
the electromagnetically-charged isosinglet $\Omega_{(12)}^-$   
decays like 
\begin{equation} 
\Omega_{(12)}^- \to \Lambda_{(2)}^0 + \widetilde{\cal P}^-_{\Theta_1}   
+ \widetilde{\cal P}^{0}_{\Theta_1}
\,,   
\end{equation} 
where the $\Omega_{(12)}^-$ has the mass larger than the $\Lambda_{(2)}^0$ mass  
due to the hypercharge-gauge boson-exchange contribution~\footnote{
The charged ${\cal P}^-_{\Theta_1}$ decays to $W^{- *} + \gamma$.}. 
The stability of the other charged isosinglet $\Omega_{(22)}^-$ is dependent on 
the choice for the $q'$ value. 
Since the sufficiently large abundance of such a stable charged particle 
has already been excluded by astrophysical observations, 
we may choose the $q'$ value to be $q' < 1$~\footnote{
The size of $q'$ would be constrained by the asymptotic safety condition 
as well as the $q$ as noted in footnote~\ref{footnote:q}. 
To avoid the Landau pole up to the Planck scale in the one-loop running of the 
$U(1)_{\BL}$ coupling $g_X$, one needs to have 
$g_X \lesssim 0.6$ for $0 < q' \le 1$, which would be reduced to 
the constraint on the $U(1)_{\BL}$ gauge boson mass, 
$m_X = g_X v_{\phi_1}/2 \lesssim {\cal O}(10^{-1}) \cdot v_{\phi_1} 
= {\cal O}(5)$ TeV for $v_{\phi_1}=50$ TeV.},  
in such a way that 
the $\Omega_{(22)}^-$ can have the mass larger than the mass of $\Omega_{(12)}^-$, 
which arises from the $U(1)_\BL$ gauge boson exchange, 
and hence is allowed to decay like 
\begin{equation}
\Omega_{(22)}^- \to \Omega_{(12)}^- + {\cal P}_{\Phi}^0
\,, 
\end{equation} 
and finally decays to $\Lambda_{(2)}^0$ as aforementioned above~\footnote{
The HC pion ${\cal P}_\Phi^{0}$ decays to diphoton, involving 
 the $\phi_1 \leftrightarrow \phi_2$ conversion twice, 
as well as the $\phi_1 \leftrightarrow \chi_{1,2}$ conversion, 
arising from 
the $y_\phi$-Yukawa coupling in Eq.(\ref{Lag}) with the $\phi$-Higgs VEV $v_{\phi_1}$.}.

Finally, consider the mass difference between $\Lambda_{(1)}^0$ 
and $\Lambda_{(2)}^0$, arising from the $U(1)_{\BL}$ gauge boson exchanges. 
It goes like $\sim \Lambda_{\rm HC} \cdot (1- 3 q')$ up to some loop factor.    
Hence the scenario will be split up to the value of $q'$: 
i) when $q'< 1/3$ or $1/3 < q' < 1$, 
either $\Lambda_{(1)}^0$ or $\Lambda_{(2)}^0$ 
decays to each of the rest, 
$\Lambda_{(2)}^0$ or $\Lambda_{(1)}^0$, along with 
the ${\cal P}_\Phi^0$ (with the $\BL$ charge $-2$);   
ii) when $q' = 1/3$, 
both of $\Lambda_{(1),(2)}^0$ are the lightest HC baryons, hence cannot decay. 
Thus, the lightest SM-singlet baryon $\Lambda^0$ ($\Lambda_{(1)}^0$ or 
$\Lambda_{(2)}^0$, or both) is most stable to be a dark matter candidate.

In the thermal history, 
the production of the $\Lambda^0$ 
has taken place in two ways;  
i) the $\Lambda^0$ can annihilate into 
other light HC hadrons such as HC pions, 
so the relic abundance would be accumulated 
by this process at around the temperature, 
$T= \Lambda_{\rm HC}={\cal O}(5-10\, {\rm TeV})$, 
through the thermal freeze-out scenario.  
However, it would not be a dominant process: 
by scaling the typical size of QCD hadron annihilating cross section, 
one gets $\langle \sigma v\rangle \sim 1/m_{\Lambda^0}^2$.  
One thus immediately finds that the freeze-out relic is negligibly small, $\Omega h^2 = {\cal O}(10^{-3})$, 
for $m_{\Lambda^0} = {\cal O}(\Lambda_{\rm HC})={\cal O}(5-10\,{\rm TeV})$; 
ii) the other possibility would be at hand, thanks to the bosonic seesaw mechanism 
as pointed out in Ref.~\cite{Ishida:2016fbp}, 
that is called the bosonic-seesaw portal process.  
In the present model, 
a source of the bosonic seesaw portal coupling can be 
generated at the $\Lambda_{\rm HC}$ scale like   
\begin{equation} 
\frac{a}{\Lambda_{\rm HC}} \overline{\Lambda}^0 
(\Theta_1^\dag \Theta_1) \Lambda^0
\,, \label{portal}
\end{equation}  
with ${\cal O}(1)$ coupling $a$. 
The bosonic seesaw, 
between the elementary Higgs doublet $H$ 
and the composite Higgs doublet $\Theta_1$,   
yields the mixing such as $\Theta_1 = y_H H_1 + \cdots \approx y_H v_{\rm EW} h_1 + \cdots$ 
for $y_H \ll 1$, where $H_1$ and $h_1$ respectively denote the 
lightest Higgs field and the physical Higgs boson field identified as  
the SM Higgs boson with the mass $m_{h_1} \simeq 125$ GeV. 
Thus the bosonic seesaw 
generates 
a Higgs portal coupling for the $\Lambda^0$ baryon: 
\begin{equation}  
a \cdot y_H \frac{v_{\rm EW}}{\Lambda_{\rm HC}}  
\left( h_1 \overline{\Lambda}^0 \Lambda^0 \right) 
\,. \label{portal:2}
\end{equation} 
As noted in~\cite{Ishida:2016fbp}, 
this coupling is still operative even  
after the particles having the mass of ${\cal O}(\Lambda_{\rm HC})$ 
decouple from the thermal equilibrium at 
around $T=\Lambda_{\rm HC}$, 
so that the $\Lambda^0$ baryon can unilaterally and nonthermally be produced 
from the SM particles through the induced-Higgs portal coupling.     
Thus this process is thought to have been dominant for the production 
in the thermal history.  
In a way done in Ref.~\cite{Ishida:2016fbp}, 
one can estimate the relic abundance to find that  
the $\Lambda^0$ baryon having the mass of ${\cal O}(5-10\,{\rm TeV})$ 
can explain the presently observed abundance of 
dark matter for $y_H \lesssim {\cal O}(10^{-5}) (\ll 1)$, provided the single component scenario.  
Note also that this smallness of the $y_H$ coupling constant is consistent with 
the bosonic seesaw formalism in Eq.(\ref{phi:seesaw}).

The $\Lambda^0$ dark matter would show up in 
the direct detection experiments such as 
the LUX and PandaX-II~\cite{Akerib:2015rjg,Akerib:2016vxi}.  
Note that in the present Case I with $q=-1/3$, 
the constituent (valence) HC fermions of the spin 1/2 $\Lambda^0 
\sim \chi_1\chi_2 \psi_{1,2}$ carry the electromagnetic charges, 
so the $\Lambda^0$ can have the electromagnetic form factors 
even though the composite state is neutral, as in the case of 
the QCD neutron. 
Among the form factors, the most stringent coupling 
to the photon arises from the magnetic moment interaction 
due to the sizable $g$ factor of ${\cal O}(1)$ generated by 
the strong dynamics. 
Such a sizable magnetic-moment portal coupling  
associated with a new strong (HC) dynamics   
has been severely constrained by direct detection experiments, 
as discussed in a context of 
some strong dynamics~\cite{Barbieri:2010mn,Banks:2010eh,Appelquist:2013ms}. 
The currently most stringent limit, derived from the LUX2016 
data~\cite{Akerib:2015rjg}, 
has been placed on the composite baryon-DM mass, 
$m_{\rm DM} > {\cal O}(10)$ 
TeV with the g factor of ${\cal O}(1)$~\footnote{  
This limit can be read off from the fourth reference in~\cite{Appelquist:2013ms} 
with a rough scaling of the upper bound of cross sections 
by a factor of 1/10 between the 2013 and 2016 data.}.

In the region satisfying $m_{\rm DM} \simeq {\cal O}(10)$ TeV,
we need the detailed analysis for the LZ and XENON1T experiments~\cite{Feng:2014uja}, including the $U(1)_{\BL}$ interaction, which is to be pursued in the future.

\subsection{Case II with $q=0$}

The HC-baryon decay chain in this case is constructed 
in a way similar to the Case I. 
First, the charged HC baryons with the higher isospin numbers  
weakly decay to the isospin partners with the lower isospin numbers: 
\begin{eqnarray} 
&& 
p_{\rm HC}^{++} \rightarrow n_{\rm HC}^+ + W^{+(*)}
\, , 
\nonumber \\ 
&& 
\Sigma_{(1),(2)}^{++} \rightarrow \Sigma_{(1),(2)}^+ + W^{+(*)} 
\rightarrow \Sigma_{(1),(2)}^0 + W^{+(*)} + W^{+(*)}
\,, \nonumber \\  
&& 
\Xi_{(11),(12),(22)}^{+} \rightarrow \Xi^0_{(11),(12),(22)} + W^{+(*)}
\,. 
\end{eqnarray}
Then the daughter HC baryons, except $\Xi_{(22)}^0$, 
 subsequently decay to 
the electromagnetically-charged isosinglet baryons $\Lambda_{(1),(2)}^+$
(having the same $\BL$ charge as those of daughters), 
plus the (off-shell) HC pions:   
\begin{eqnarray} 
&& 
n_{\rm HC}^{+} \rightarrow \Lambda_{(2)}^+ + {\cal P}_{\Theta_2}^0
\,, \nonumber \\  
&& 
\Sigma_{(1),(2)}^0 \rightarrow \Lambda_{(1),(2)}^+ + {\cal P}_{a_0^{\rm HC}}^0 
+ {\cal P}_{a_0^{\rm HC}}^-, 
\nonumber \\  
&& 
\Xi_{(11),(22)}^0 \rightarrow \Lambda_{(1),(2)}^+ + \widetilde{\cal P}_{\Theta_1, \Theta_2}^-
\,. 
\end{eqnarray}   
The rest, $\Xi^0_{(22)}$, decays to the electromagnetically neutral-isosinglet 
$\Omega_{(22)}^0$ along with the isospin-doublet HC pion ${\cal P}_{\Theta_1}^0$.

The stability of the singly-charged $\Lambda_{(1), (2)}^+$ depends on the $\BL$ charge value $q'$. 
To avoid a stable charged baryon, as done in the Case I, 
we may take $q' \ge 1/3$, so as to allow  
the decay channel of the $\Lambda_{(1)}^+$ to the $\Lambda_{(2)}^+$: 
\begin{equation} 
\Lambda_{(1)}^+ \rightarrow \Lambda_{(2)}^+ + {\widetilde {\cal P}}_\Phi^{0}
\,. 
\end{equation} 
Note that we have the mass difference between $\Lambda_{(1)}^+$ and 
$\Lambda_{(2)}^+$, $\Delta m_{(1)-(2)}\propto (3 q' -1)>0$, according to the $\BL$ charge. 
$\Lambda_{(2)}^+$ can decay 
to the neutral $\Omega_{(12)}^0$, 
by emitting the (off-shell) two isospin-doublet HC pions ${\cal P}_{\Theta_1}^+$ 
and ${\cal P}_{\Theta_1}^0$.

For the remaining $\Omega_{(12),(22)}^0$ baryons, the stability 
again depends on the value of the $\BL$ charge, $q'$: 
when $q'\neq 1$ is satisfied, either $\Omega_{(12)}^0$ or $\Omega_{(22)}^0$ 
can decay to either of the rest, along with the HC pion ${\cal P}_{\Phi}^0$ 
or ${\widetilde {\cal P}}_\Phi^{0}$. 
In the case of $q'=1$, these two baryons are degenerate so that 
both two can be DM candidates. 
Thus, the lightest $\Omega^0$ ($\Omega_{(12)}^0$ or $\Omega_{(22)}^0$, or both)  
becomes stable when the $\BL$ charge is taken as $q'\ge 1/3$.

The thermal history of the $\Omega^0$ is 
the same as the $\Lambda^0$ in the Case I:  
the relic abundance has dominantly been produced 
by the bosonic-seesaw portal process with the portal coupling 
as in Eq.(\ref{portal:2}) replacing $\Lambda^0$ with $\Omega^0$. 
The desired amount of the abundance 
can thus be accumulated consistently with 
the bosonic seesaw mechanism with the small coupling 
$y_H \lesssim {\cal O}(10^{-5})$~\cite{Ishida:2016fbp}.

As to the discovery sensitivity in 
direct detection experiments, 
it is drastically different from the Case I: 
since the constituent HC fermions of the spin 1/2 
$\Omega^0$ baryon do not have 
the electromagnetic charges, 
the magnetic moment cannot be generated, 
so the $\Omega^0$ DM is free from the severe constraint 
on the photon portal process in direct detection experiments 
as discussed in Refs.~\cite{Barbieri:2010mn,Banks:2010eh,Appelquist:2013ms}.    
The most dominant source 
then turns out to be the $U(1)_{\BL}$-gauge boson portal process. 
(As noted in Ref.~\cite{Ishida:2016fbp}, 
the bosonic-seesaw portal coupling yields 
tiny spin-independent nucleon-dark matter scattering cross section, 
to be negligible compared to the $\BL$ portal contribution.)

Indeed,  
the $\Omega^0$-$\Omega^0$-$U(1)_{\BL}$ gauge boson 
interaction would be sizable enough to get sensitive at the direct detection 
experiment. 
This feature is in contrast to the literature~\cite{Ishida:2016fbp}, in which 
the $U(1)_{\BL}$ gauge has not been introduced.    
As done in the effective operator analysis, say, in Ref.~\cite{Agrawal:2010fh}, 
the four-fermion coupling $(g_X^2/m_X^2)=1/v_{\phi_1}^2$ induced from 
the $U(1)_{\BL}$ gauge-boson $(X)$ exchange between the $\Omega^0$ and 
nucleon currents is constrained by the exclusion limits provided by 
the detection experiments. 
The currently most stringent limit from the LUX2016~\cite{Akerib:2016vxi} 
thus constrains the $\phi$-Higgs VEV $v_{\phi_1}$. 
When $q'=1$ is taken for a benchmark value, we find 
the lower bound,  
$v_{\phi_1} > 5.1 (4.3)$ TeV,for the $\Omega^0$ 
DM mass $m_{\Omega^0}=5(10)$ TeV.   
The prospected future detection experiments such as the XENON1T and LZ~\cite{Feng:2014uja} 
will give more severe limits, to constrain the parameter space in the model.

\section{Conclusion}\label{Sec5}  
In the model presented here, the dynamical scalegenesis has 
successfully generated masses of the standard-model particle and active neutrino as well as explained 
the dark matter, by the multiple seesaw mechanisms induced from the new strong dynamics of the hypercolor. 
We have predicted a number of hypercolor hadrons, 
the $\BL$ Higgs, gauge bosons and three heavy right-handed neutrinos, around the order of a few or tens of TeV scale.

Some of hypercolor baryons, with the mass on the order of 
the hypercolor scale $\Lambda_{\rm HC}$, say, ${\cal O}(5-10)$ TeV,  
can be stabilized due to 
the hypercolor-baryon number and the $\BL$ charge. 
Two classes of dark matter candidates have been discussed 
by splitting the model in terms of the hypercharge parameter ($q$).  
Note that the success of the dynamical scalegenesis is irrespective to 
the $q$ value.    
Those two classes are shown to have different sensitivities 
to dark-matter detection experiments: 
one scenario implies that 
the dominant source to measure the dark matter 
is provided by the potentially 
large magnetic-moment form factor generated by 
the strongly coupled hypercolor dynamics (called the Case I), 
while the other by the $\BL$ gauge-boson portal coupling (Case II).   
Future-planned detection experiments such as the XENON1T and LZ 
would make it possible 
to clearly verify which scenario would be favored.

The model can also be tested by the collider 
signatures of those new particles, as well as the searches for 
the dark matter. 
In particular, the lightest hypercolor hadrons, say, the hypercolor pions 
would show up at the LHC with distinct signals, 
as addressed in the literature~\cite{Ishida:2016ogu}, 
so would be a smoking gun of this model.

More details on the phenomenological analyses, including 
collider study of the hypercolor hadrons in correlation with 
the $U(1)_\BL$ gauge boson and flavor physics induced by the 
couplings to the heavy Higgs $(H_2)$, will be pursued in the future.   
\\

\acknowledgments 
We are grateful to David Schaich for useful comments 
on direct detection experiments for baryonic 
composite dark matters.  
This work was supported in part by 
the JSPS Grant-in-Aid for Young Scientists (B) \#15K17645 (S.M.).

\appendix

\section{The realization of EW and $U(1)_{\BL}$ symmetry breaking}\label{App1} 
\label{appendix:2}

In this Appendix we analyze the potential terms regarding 
the realization of the EWSB and the $U(1)_{\BL}$ gauge symmetry breaking.

We employ an effective potential at the $\Lambda_{\rm HC}$ scale 
including terms in Eq.(\ref{Lag}) 
and the HC-induced terms, 
\begin{eqnarray} 
V &=& \lambda_H (H^\dag H)^2 + \lambda_\phi (|\phi|^2)^2 
+ \kappa_\phi |\phi|^2 (H^\dag H) 
+ y_H \Lambda_{\rm HC}^2 \left( \Theta_1^\dag H +{\rm h.c.}  \right) 
+ y_\phi \Lambda_{\rm HC}^2 \left( \Phi^* \phi + {\rm h.c.} \right) 
\nonumber \\ 
&& 
+ m_{{\cal M}}^2 {\rm tr}[{\cal M}^\dag {\cal M}] 
+ \lambda_1 {\rm tr}[({\cal M}^\dag {\cal M})^2] 
+ \lambda_2 ({\rm tr}[{\cal M}^\dag {\cal M}])^2 
\,, 
\label{potential}
\end{eqnarray} 
where ${\cal M}={\cal S} + i {\cal P}$ denotes the 
composite scalar field of the $4 \times 4$ matrix form, transforming bifundamentally 
under the $U(4)_{F_L} \times U(4)_{F_R}$ symmetry, ${\cal M} \to g_L \cdot {\cal M} \cdot 
g_R^\dag$ with $g_{L/R} \in U(4)_{F_{L/R}}$. 
The ${\cal M}$ is expanded with respect to the $U(4)$ generators $T^a$ ($a=0,\cdots, 15$),  
normalized by ${\rm tr}[T^aT^b]=\delta^{ab}/2$ with $T^0 = 1/2\sqrt{2}\cdot 1_{4\times 4}$, 
as ${\cal M}=\sum_a {\cal M}^a T^a$.    
In terms of the ${\cal S}^a$, the composite Higgs doublet $\Theta_1$ and the composite 
$\BL$ Higgs $\Phi$ are parametrized as 
\begin{eqnarray} 
\Theta_1 
&=& 
\left(
\begin{array}{c} 
\Theta_1^+ \\ 
\Theta_1^0 
\end{array}
\right) 
= 
 \left(
\begin{array}{c} 
\frac{{\cal S}^4 - i {\cal S}^5}{\sqrt{2}} \\ 
\frac{{\cal S}^6 - i {\cal S}^7}{\sqrt{2}} 
\end{array}
\right) 
\,, \nonumber \\ 
\Phi &=& \frac{{\cal S}^{13} - i {\cal S}^{14}}{\sqrt{2}}
\,. 
\end{eqnarray}

Since the $y_H$ and $y_\phi$ couplings are assumed to be 
small ($y_{H,\phi} \ll 1$), 
the potential in Eq.(\ref{potential}) possesses 
the approximate ``chiral" $U(4)_{F_L} \times U(4)_{F_R}$ 
symmetry reflected in the ${\cal M}$ sector. 
Matching with the underlying vectorlike dynamics of the HC, 
we choose the VEV of ${\cal S}$, $\langle {\cal S} \rangle = {\cal S}^0/2\sqrt{2} \cdot 1_{4 \times 4} 
= v/2\sqrt{2} \cdot 1_{4 \times 4}$, 
to realize the spontaneous breaking pattern $U(4)_{F_L} \times U(4)_{F_R} 
\to SU(4)_{F_V} \times U(1)_{F_V}$, with the $U(1)_{F_A}$ anomaly 
in the underlying HC dynamics taken into account. 
(The state ${\cal S}^0$ corresponds to a linear combination of $f_0^{\rm HC}$, the third-adjoint component of $a_0^{\rm HC}$, 
$\varphi_1$ and $\varphi_2$ in Table~\ref{tab:3}.)
The VEV $v$ is equivalent to the HC pion decay constant $f_{\cal P}$, 
which can be related to the $\Lambda_{\rm HC}$ scale as 
$f_{\cal P} \simeq  \Lambda_{\rm HC}/(4\pi) = {\cal O}(1)$ TeV 
for $\Lambda_{\rm HC}={\cal O}(5-10\,{\rm TeV})$.  
The stationary condition for the $v$ is then derived from Eq.(\ref{potential})
to be 
\begin{equation} 
 v \left( m_{{\cal M}}^2 + \left( \frac{\lambda_1}{4} + \lambda_2 \right) v^2 \right) = 0 
 \,, 
\end{equation}
so that we have the VEV $v^2 = - m_{\cal M}^2/(\lambda_1/4+ \lambda_2)$.

The physical ${\cal S}^0$ scalar mass arises by expanding the potential around 
the VEV $v$, to be 
\begin{equation} 
 m_{{\cal S}^0} = \sqrt{2 (\lambda_1/4+ \lambda_2)} v 
 \,. 
\end{equation}

As clearly seen from the potential form in Eq.(\ref{potential}), 
one can always choose the vacuum where 
the VEVs of composite scalars are zero (i.e. trivial solutions for the stationary conditions), 
except for the $\Theta_1$ and 
$\Phi$ having the quadratic mixing terms with the elementary $H$ and $\phi$. 
Therefore, we can extract only the $\Theta_1$ and $\Phi$ scalars from the ${\cal M}$ 
matrix in the potential Eq.(\ref{potential}), and derive the effective potential terms 
relevant to discussion on the EWSB and the $U(1)_{\BL}$ breaking: 
\begin{eqnarray} 
V_{\rm eff} &=& \lambda_H (H^\dag H)^2 + \lambda_\phi (|\phi|^2)^2 
+ \kappa_\phi |\phi|^2 (H^\dag H) 
+ y_H \Lambda_{\rm HC}^2 \left( \Theta_1^\dag H +{\rm h.c.}  \right) 
+ y_\phi \Lambda_{\rm HC}^2 \left( \Phi^* \phi + {\rm h.c.} \right) 
\nonumber \\ 
&& 
+ m^2_{\cal S} \left[ (\Theta_1^\dag \Theta_1) + |\Phi|^2 \right] 
+ \lambda_{\cal S} (\Theta_1^\dag \Theta_1 + |\Phi|^2)^2
\,, \label{potential:eff}
\end{eqnarray} 
where 
$
 m^2_{\cal S} = (3\lambda_1/8) v^2 
 (\simeq 3\Lambda_{\rm HC}^2/16 )
$ 
and 
$ 
\lambda_{\cal S} = \lambda_1/2 + \lambda_2 
$. 
Note the degenerate mass and quartic coupling terms for $\Theta_1$ and $\Phi$, 
reflecting the approximate ``chiral" $SU(4)_{F_L} \times SU(4)_{F_R}$ symmetry.

Solving the quadratic mixing terms for $\Theta_1$-$H$ and $\Phi$-$\phi$ of the 
seesaw form in Eq.(\ref{potential:eff}), 
to the leading order of expansion in $y_H$ and $y_\phi$,  
one finds the mass eigenstate fields $(H_1, H_2)$ and $(\phi_1, \phi_2)$ 
related to the original fields $(H,\Theta)$ and $(\phi, \Phi)$ as 
\begin{eqnarray} 
 \left( 
\begin{array}{c} 
H_1 \\ 
H_2 
\end{array}
\right) 
&\simeq &
\left(
\begin{array}{cc} 
1 & - y_H \\ 
y_H & 1 
\end{array}
\right) 
 \left( 
\begin{array}{c} 
H \\ 
\Theta_1 
\end{array}
\right)
\,, \nonumber \\ 
 \left( 
\begin{array}{c} 
\phi_1 \\ 
\phi_2 
\end{array}
\right) 
&\simeq &
\left(
\begin{array}{cc} 
1 & - y_\phi \\ 
y_\phi & 1 
\end{array}
\right) 
 \left( 
\begin{array}{c} 
\phi \\ 
\Phi 
\end{array}
\right)
\,,
\end{eqnarray}
 with the mass eigenvalues 
\begin{eqnarray}
m_{H_1}^2 
&\simeq& - y_H^2 \frac{\Lambda_{\rm HC}^4}{m_{\cal S}^2} 
\left(\simeq - \frac{16}{3} y_H^2 \Lambda_{\rm HC}^2 \right) 
\,, \nonumber \\ 
m_{H_2}^2 &\simeq& 
m_{\cal S}^2 \left( 
\simeq \frac{3}{16} \Lambda_{\rm HC}^2 \right) 
\,, \nonumber \\  
m_{\phi_1}^2 &\simeq& 
- y_\phi^2 \frac{\Lambda_{\rm HC}^4}{m_{\cal S}^2} 
\left(\simeq - \frac{16}{3} y_\phi^2 \Lambda_{\rm HC}^2 \right) 
\,, \nonumber \\ 
m_{\phi_2}^2 &\simeq& 
m_{\cal S}^2 \left( 
\simeq \frac{3}{16} \Lambda_{\rm HC}^2 \right) 
\,. 
\end{eqnarray} 
Plugging these expressions into the effective potential and rewriting 
the terms in terms of the mass eigenstate fields, 
one finds the stationary conditions under the assumption that the $H_2$ and $\phi_2$ 
do not develop the VEVs: 
\begin{eqnarray} 
- m_{H_1}^2 &\simeq& \frac{1}{2} \lambda_{\cal S} \left(y_H^2 v_1^2 + y_\phi^2 v_{\phi_1}^2 \right) 
\,, \nonumber \\ 
- m_\phi^2 &\simeq & 4 \lambda_\phi v_{\phi_1}^2
\,, \nonumber \\ 
- \kappa_\phi v_{\phi_1}^2 &\simeq& \lambda_H v_1^2 
\,, \label{condi}
\end{eqnarray}
where $v_1$ and $v_{\phi_1}$ stand for the VEVs of $H_1$ and $\phi_1$, respectively, and 
the last condition has come from the vacuum assumption. 
Thus, we realize the EWSB $(v_1 (\neq 0) \simeq 246\,{\rm GeV})$ 
and $U(1)_{\BL}$ gauge symmetry breaking $(v_{\phi_1} \neq 0)$.

Taking into account the stationary conditions in Eq.(\ref{condi}) 
and expanding the $H_1$ and $H_2$ around those VEVs 
as $H_1 = \frac{1}{\sqrt{2}} (0, v_1 + h_1)^T$, $H_2=\frac{1}{\sqrt{2}}(0,h_2)^T$, and redefining as 
$\phi_1 \to \frac{1}{\sqrt{2}} (v_{\phi_1} + \phi_1)$ 
and $\phi_2 \to \frac{1}{\sqrt{2}} \phi_2$,  
one can find the physical masses in the effective potential, 
\begin{eqnarray} 
m_{h_1} &\simeq& \sqrt{2 \lambda_H} v_1 (\simeq 125\,{\rm GeV}) 
\,, \nonumber \\ 
m_{h_2} & \simeq & m_{\phi_2} \simeq m_{\cal S} \left( \simeq \frac{\sqrt{3}}{4} \Lambda_{\rm HC} \right)
\,, \nonumber \\ 
m_{\phi_1} &\simeq & 2 \sqrt{2 \lambda_\phi} v_{\phi_1} 
\,.  
\end{eqnarray}


\begin{thebibliography}{99} 
%
\bibitem{Aad:2012tfa}
  G.~Aad {\it et al.} [ATLAS Collaboration],
  Phys.\ Lett.\ B {\bf 716} (2012) 1
  [arXiv:1207.7214 [hep-ex]].

%
\bibitem{Chatrchyan:2013lba}
  S.~Chatrchyan {\it et al.} [CMS Collaboration],
  JHEP {\bf 1306} (2013) 081
  [arXiv:1303.4571 [hep-ex]].

%
\bibitem{Shaposhnikov:2009pv}
  M.~Shaposhnikov and C.~Wetterich,
  Phys.\ Lett.\ B {\bf 683} (2010) 196
  [arXiv:0912.0208 [hep-th]]; 
  C.~Wetterich and M.~Yamada,
  arXiv:1612.03069 [hep-th].


%
\bibitem{Helmboldt:2016mpi}
  A.~J.~Helmboldt, P.~Humbert, M.~Lindner and J.~Smirnov,
  arXiv:1603.03603 [hep-ph].


%
\bibitem{Coleman:1973jx} 
  S.~R.~Coleman and E.~J.~Weinberg,
  Phys.\ Rev.\ D {\bf 7}, 1888 (1973).





\bibitem{B-L}
  S.~Iso, N.~Okada and Y.~Orikasa,
  Phys.\ Rev.\ D {\bf 80}, 115007 (2009)
  [arXiv:0909.0128 [hep-ph]]; 
%
  S.~Iso, N.~Okada and Y.~Orikasa,
  Phys.\ Lett.\ B {\bf 676}, 81 (2009)
  [arXiv:0902.4050 [hep-ph]]; 
%
  N.~Okada and Y.~Orikasa,
  Phys.\ Rev.\ D {\bf 85}, 115006 (2012)
  [arXiv:1202.1405 [hep-ph]]; 
%
  S.~Iso and Y.~Orikasa,
  PTEP {\bf 2013}, 023B08 (2013)
  [arXiv:1210.2848 [hep-ph]]; 
%
  I.~Oda,
  Phys.\ Lett.\ B {\bf 724}, 160 (2013)
  [arXiv:1305.0884 [hep-ph]]; 
%
  V.~V.~Khoze and G.~Ro,
  JHEP {\bf 1310}, 075 (2013)
  [arXiv:1307.3764 [hep-ph]]; 
%
  M.~Hashimoto, S.~Iso and Y.~Orikasa,
  Phys.\ Rev.\ D {\bf 89}, no. 5, 056010 (2014)
  [arXiv:1401.5944 [hep-ph]].  
%






\bibitem{nu:sumaary}
See, for instance, 
  M.~Lindner, S.~Schmidt and J.~Smirnov,
  JHEP {\bf 1410} (2014) 177
  [arXiv:1405.6204 [hep-ph]],  
which exemplifies 
models to realize the neutrino masses from a conformal-EW symmetry-breaking 
with or without introducing the B-L gauge. 


\bibitem{Okada:2014nea}
  H.~Okada and Y.~Orikasa,
  Phys.\ Lett.\ B {\bf 760} (2016) 558
  [arXiv:1412.3616 [hep-ph]]; 
%
  J.~Guo, Z.~Kang, P.~Ko and Y.~Orikasa,
  Phys.\ Rev.\ D {\bf 91}, no. 11, 115017 (2015)
  [arXiv:1502.00508 [hep-ph]].
%
%






%
\bibitem{Haba:2015lka}
  N.~Haba, H.~Ishida, N.~Okada and Y.~Yamaguchi,
  Phys.\ Lett.\ B {\bf 754} (2016) 349
  [arXiv:1508.06828 [hep-ph]].




\bibitem{BSS}
  X.~Calmet,
  Eur.\ Phys.\ J.\ C {\bf 28} (2003) 451
  [hep-ph/0206091]; 
  H.~D.~Kim,
  Phys.\ Rev.\ D {\bf 72} (2005) 055015
  [hep-ph/0501059]; 
  N.~Haba, N.~Kitazawa and N.~Okada,
  Acta Phys.\ Polon.\ B {\bf 40} (2009) 67
  [hep-ph/0504279]; 
  O.~Antipin, M.~Redi and A.~Strumia,
  JHEP {\bf 1501} (2015) 157
  [arXiv:1410.1817 [hep-ph]].

%
\bibitem{Haba:2015qbz}
  N.~Haba, H.~Ishida, N.~Kitazawa and Y.~Yamaguchi,
  Phys.\ Lett.\ B {\bf 755} (2016) 439
  [arXiv:1512.05061 [hep-ph]].
%
\bibitem{Ishida:2016ogu} 
  H.~Ishida, S.~Matsuzaki and Y.~Yamaguchi,
  Phys.\ Rev.\ D {\bf 94}, no. 9, 095011 (2016)
  [arXiv:1604.07712 [hep-ph]].
%
\bibitem{Ishida:2016fbp}
  H.~Ishida, S.~Matsuzaki and Y.~Yamaguchi,
  arXiv:1610.07137 [hep-ph].
%
\bibitem{Hid-strong-EWSB}
  F.~Wilczek,
  Int.\ J.\ Mod.\ Phys.\ A {\bf 23} (2008) 1791
   [Eur.\ Phys.\ J.\ C {\bf 59} (2009) 185]
  [arXiv:0708.4236 [hep-ph]]; 
  T.~Hur and P.~Ko,
  Phys.\ Rev.\ Lett.\  {\bf 106} (2011) 141802
  [arXiv:1103.2571 [hep-ph]]; 
  M.~Holthausen, J.~Kubo, K.~S.~Lim and M.~Lindner,
  JHEP {\bf 1312} (2013) 076
  [arXiv:1310.4423 [hep-ph]]; 
  J.~Kubo, K.~S.~Lim and M.~Lindner,
  Phys.\ Rev.\ Lett.\  {\bf 113} (2014) 091604
  [arXiv:1403.4262 [hep-ph]].
%
\bibitem{seesaw} 
  P.~Minkowski,
  Phys.\ Lett.\ B {\bf 67}, 421 (1977); 
%
T. Yanagida, in Proceedings of the Workshop on Unified Theory and Baryon Number of the Universe, edited by O. Sawada and A. Sugamoto (KEK, Tsukuba, Ibaraki 305- 0801 Japan, 1979) p. 95; 
%
  T.~Yanagida,
  Prog.\ Theor.\ Phys.\  {\bf 64}, 1103 (1980); 
%
  M. Gell-Mann, P. Ramond, and R. Slansky, in Supergravity, edited by P. van Niewwenhuizen and D. Freedman (North Holland, Amsterdam, 1979)
  [arXiv:1306.4669 [hep-th]]; 
%
P.~Ramond, 
in {\em Talk given at the Sanibel Symposium}, 
Palm Coast, Fla., Feb.~25-Mar.~2, 1979, preprint CALT-68-709
(retroprinted as hep-ph/9809459); 
%
S.~L.~Glashow,
in {\em Proc. of the Carg\'ese  Summer Institute on Quarks and Leptons},
Carg\'ese, July 9-29, 1979, 
eds. M.~L\'evy et. al, , (Plenum, 1980, New York), p707; 
%
%
%
  R.~N.~Mohapatra and G.~Senjanovic,
  Phys.\ Rev.\ Lett.\  {\bf 44} (1980) 912; 
  J.~Schechter and J.~W.~F.~Valle,
  Phys.\ Rev.\ D {\bf 22} (1980) 2227; 
  J.~Schechter and J.~W.~F.~Valle,
  Phys.\ Rev.\ D {\bf 25} (1982) 774.

\bibitem{Akerib:2015rjg} 
  D.~S.~Akerib {\it et al.} [LUX Collaboration],
  Phys.\ Rev.\ Lett.\  {\bf 116}, no. 16, 161301 (2016) 
  [arXiv:1512.03506 [astro-ph.CO]].

\bibitem{Akerib:2016vxi} 
  D.~S.~Akerib {\it et al.},
   arXiv:1608.07648  [astro-ph.CO]; 
  A.~Tan {\it et al.}    [PandaX-II Collaboration],
   Phys.\ Rev.\ Lett.\  {\bf 117}, no. 12, 121303 (2016) 
  [arXiv:1607.07400 [hep-ex]].





\bibitem{Barbieri:2010mn} 
  R.~Barbieri, S.~Rychkov and R.~Torre,
  Phys.\ Lett.\ B {\bf 688}, 212 (2010)
  [arXiv:1001.3149 [hep-ph]].

\bibitem{Banks:2010eh} 
  T.~Banks, J.~F.~Fortin and S.~Thomas,
  arXiv:1007.5515 [hep-ph].

\bibitem{Appelquist:2013ms} 
  T.~Appelquist {\it et al.} [Lattice Strong Dynamics (LSD) Collaboration],
  Phys.\ Rev.\ D {\bf 88}, no. 1, 014502 (2013)
  [arXiv:1301.1693 [hep-ph]], 
and the fourth reference of~\cite{BSS}. 



\bibitem{Feng:2014uja} 
  J.~L.~Feng {\it et al.},
  arXiv:1401.6085 [hep-ex].


\bibitem{Agrawal:2010fh} 
  P.~Agrawal, Z.~Chacko, C.~Kilic and R.~K.~Mishra,
  arXiv:1003.1912 [hep-ph]; 
  M.~Cirelli, E.~Del Nobile and P.~Panci,
  JCAP {\bf 1310}, 019 (2013)
  [arXiv:1307.5955 [hep-ph]].





\end{thebibliography}
\end{document}